\documentclass[aps,showkeys,groupedaddress]{revtex4}

\usepackage{epsfig}
\usepackage{epsf}

\begin{document}


%
%

\title{Phase Transition and Monopoles Densities in a\\
Nearest Neighbors Two-Dimensional Spin Ice Model}

\author{C. W. MORAIS, D. N. DE FREITAS, A. L. MOTA\footnote{motaal@ufsj.edu.br}  ~and E. C. BASTONE}

\affiliation{Departamento de Ci\^{e}ncias Naturais, Universidade Federal de S\~{a}o Jo\~{a}o del-Rei, \\ Campus Dom Bosco, Pra\c{c}a Dom Helv\'{e}cio, 74, CEP 36301-160,\\ S\~ao Jo\~ao del-Rei, Minas Gerais, Brazil }



\begin{abstract}
In this work, we show that, due to the alternating orientation of the spins in the ground state of the artificial square spin ice, the influence of a set of spins at a certain distance of a reference spin decreases faster than the expected result for the long range dipolar interaction, justifying the use of the nearest neighbor two dimensional square spin ice model \cite{Xie2015} as an effective model. Using an extension of the model presented in ref. \cite{Xie2015}, considering the influence of the eight nearest neighbors of each spin on the lattice, we analyze the thermodynamics of the model and study the monopoles and string densities dependence as a function of the temperature.
\end{abstract}


\maketitle

\section{Introduction}

Spin ice are ferromagnetic frustrated rare-earth materials with pyrochlore structure \cite{Harris1997,Ramirez1999,Bramwell2001b,Ramirez2002}. The rare earth ions are located at the corners of tetrahedrons, and due to the anisotropy their magnetic moments can point only in the two possible orientations in the direction that connects the corner to the center of the tetrahedron - acting like Ising spin variables. The exchange and dipolar interactions between the magnetic moments, as well as the geometry of the structure, impose the so called ice rule - at the ground state, two of the spins located at the corners point inward and the two other spins point outward each tetrahedron. Spin ice materials present also excitations (violation of this 2-in, 2-out ice rule) that behaves like magnetic monopoles \cite{Castelnovo2008,Morris2009}, rising both theoretical \cite{Snyder2001,Melko2004,Isakov2005,Jaubert2009,Udagawa2015} and experimental \cite{Bramwell2001,Fukazawa2002,Mirebeau2005,Kadowaki2009,Bramwell2009,Giblin2011} interest in the study of the system.

As the spin-orbit coupling is large in rare-earth ions, the interactions on a spin ice structure is dominated by the exchange interaction \cite{Hertog2000,Bramwell2001}. Nevertheless, the dipolar long range interactions are comparable in magnitude, and also plays a significant role in the description of the system, so spin ice is usually described by an effective pyrochlore Ising model \cite{Bramwell2004} that includes both exchange and dipolar interactions. In a simplified version of the the spin ice model, the dipolar coefficient is set to zero, and the exchange interaction is considered only between nearest neighbors. This simplified version is referred as the nearest neighbor spin ice model (nnSI) \cite{Hertog2000,Bramwell2001,Isakov2005,Guruciaga2014,Ferreyra2016}, and was employed to the qualitative description of several properties of spin ice materials.

Belonging to a different class of materials, but with some similar properties, two-dimensional spin ices are artificial structures, litographically assembled, constituted by elongated permalloy nano-islands \cite{Wang2006,Moller2006}. The magnetic moments of the nano-islands are, due to the islands shape anisotropy, constricted to point on the direction of the longest axis \cite{Wysin2012}. Again, magnetic moments behave as Ising spin variables, being referred only as spins. If the geometry of the array is a two-dimensional square lattice, in which the islands alternate in orientation along the $x$ and $y$ axis, as depicted in fig. \ref{sjk}(a), the system is denominated a two-dimensional square spin ice.
Exchange interactions are not present, due to the dimensions of the nano-islands and lattice spacing, providing a classical character to the phenomena here studied, but the dipolar long range interactions is still present and it is responsible for the two-dimensional version of the ice rule - at the ground state, in each vertex of the array of nano-islands, two spins point in the direction of the vertex and two others point in the opposite direction. Also these two-dimensional spin ice systems are subject of great interest, and several studies on the thermodynamics and dynamics of the system and its excitations - the magnetic monopoles - are conducted \cite{Qi2008,Mellado2010,Budrikis2010,Ladak2010,Mengotti2011,Budrikis2012,Kapaklis2012,Zhang2013,Kapaklis2014}.

In a recent work, Xie, Du, Yan and Liu \cite{Xie2015} used a nearest neighbors model to study phase transitions in a square two-dimensional spin ice, in the context of the conserved monopoles algorithm \cite{Borzi2013}, to analyze the thermodynamics of monopoles in an ensemble with fixed monopole density. The authors consider first a model Hamiltonian presenting the exchange interactions between the nearest neighbors spin pairs as well as the long range dipolar interaction. This model Hamiltonian can represent both the artificial two-dimensional square spin ice, lithographically constructed, and also the square spin ice mapping obtained from the projection of the three-dimensional tetrahedral lattice of the pyrochlore spin ice on a plane \cite{Mol2010}. Since spin frustration comes from the nearest neighbors interactions, this Hamiltonian is then reduced to an effective nearest neighbors Hamiltonian given by
\begin{equation}
H = \frac{J_{eff}}{2} \sum_{<j,k>} S_j S_k,
\end{equation}
where the sum runs over the nearest neighbors spin pairs and $J_{eff}$ is the effective coupling that include both exchange and dipolar interactions and assume different values for the two different alignments of the pair of spins (perpendicular or parallel). Following the denomination employed for the nearest neighbors spin ice model, we will refer to this model as the nearest neighbors two-dimensional spin ice model (nn2DSI).

In the present work we will use an extension of the nn2DSI model to study the thermodynamics of the two dimensional square spin ice in the canonical ensemble and compare its results to those obtained by considering the long range character of the dipolar interaction, presented in references \cite{Silva2012,Li2013}. There, the authors used Monte Carlo simulations to develop a study of the thermodynamics of the model, including the interaction between all the pairs of spins on finite lattices, with periodic boundary conditions treated by the Ewald summation technique \cite{Zuowei2001} or with open boundary conditions. These approaches, of course, made the results more precise, but with the cost of increasing the computational efforts. 

The extension of the model we present here will consider a set of eight spins as the nearest neighbors of a given spin, in contrast with the six nearest neighbors of the nn2DSI model \cite{Xie2015}, for the reasons we will present in section \ref{sri}. Also in section \ref{sri}, we will show that the major influence on a reference spin comes from this set of eight neighbors, in comparison to the influence of all the other spins on the lattice, justifying the use of the nn2DSI model as an effective model for two-dimensional spin ice. In section \ref{nn2DSI} we define the model Hamiltonian to the extension of the nn2DSI model employed here. In section \ref{nr} we present the numerical results of our Monte-Carlo simulations, including the phase transition from the ordered to the disordered phase and the behavior of the density of monopoles and strings on the lattice as a function of the temperature. Finally, we present our conclusions in section \ref{conclusions}.

\section{The nearest neighbors approach} \label{sri}

\begin{figure}[ht]
\begin{center}
\includegraphics[angle=0,width=0.35\columnwidth]{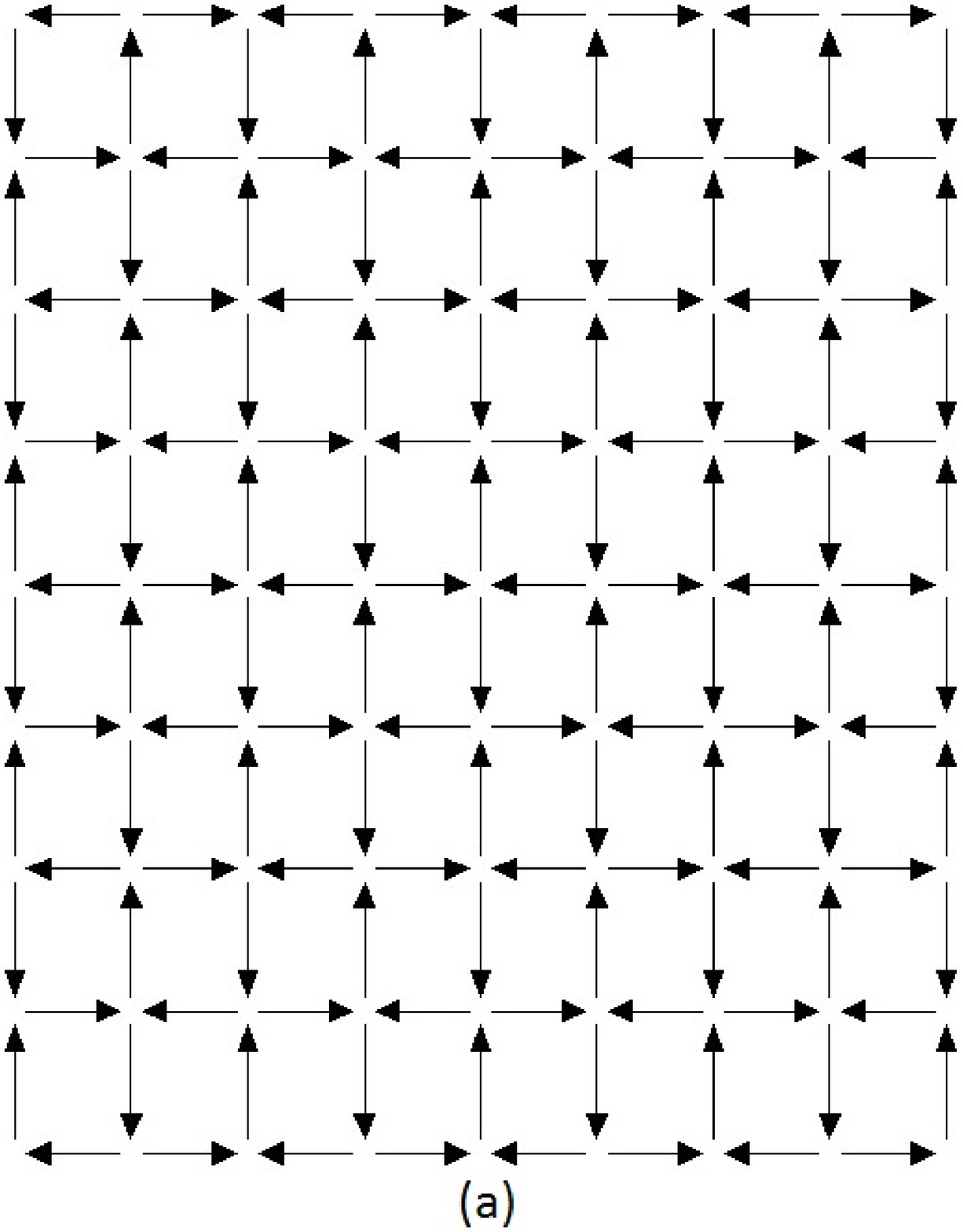} \includegraphics[angle=0,width=0.35\columnwidth]{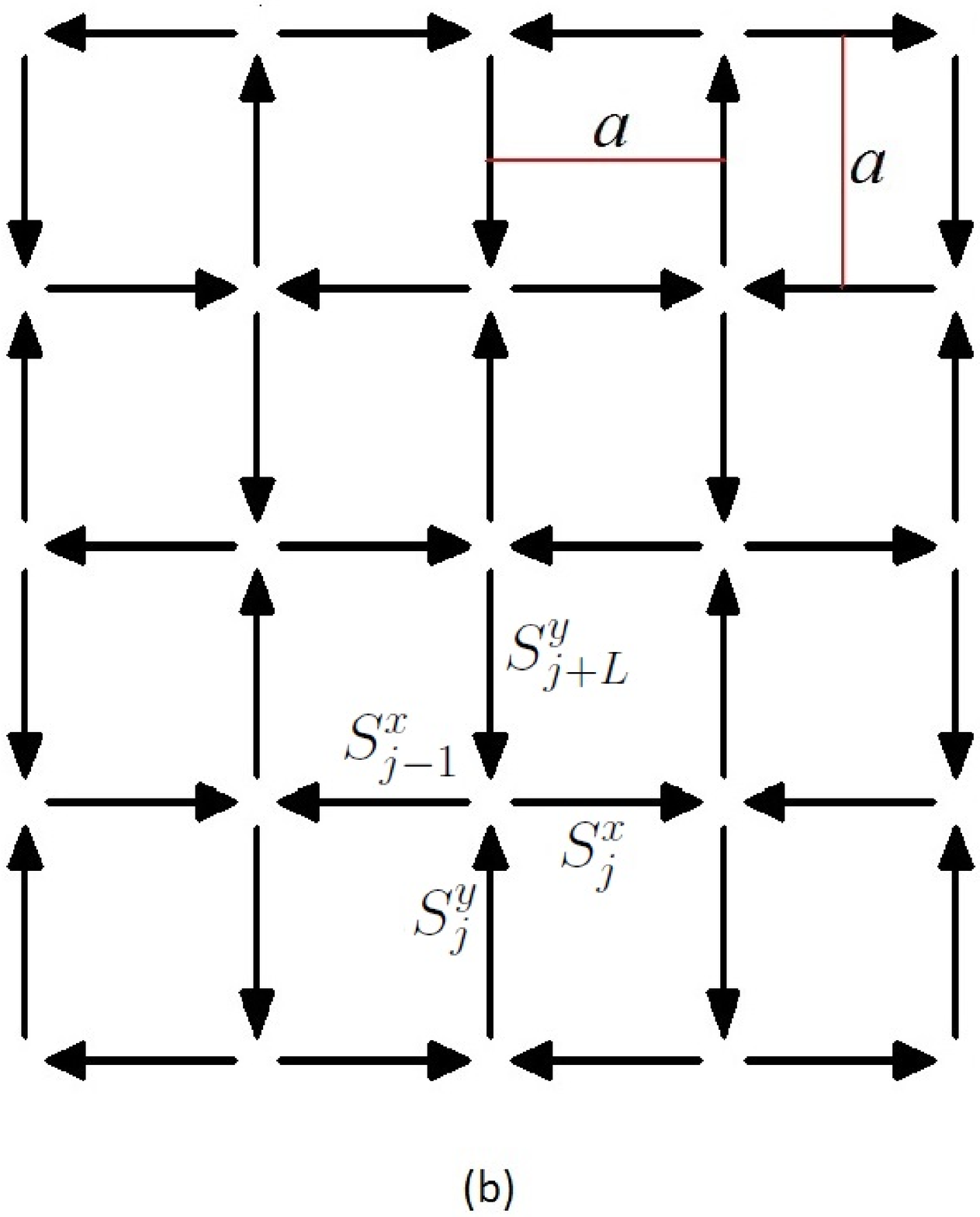}
\end{center}
\caption{(a) Square spin ice ground state. (b) The spins adjacent to a vertex $j$. The lattice spacing is denoted by $a$.}
\label{sjk}
\end{figure}

The dipolar interaction between the permalloy nano-islands on the artificial two-dimensional spin ice model can be described by the dipolar Hamiltonian given by
\begin{equation}
H = D a^3 \sum_{jk} \Big( \frac{\vec{S}_i \cdot \vec{S}_j}{r^3_{jk}} - 3 \frac{(\vec{S}_i \cdot \vec{r}_{jk})(\vec{S}_j \cdot \vec{r}_{jk})}{r^5_{jk}}\Big), \label{hamiltonian}
\end{equation}
where $D=\frac{\mu_0 \mu}{4 \pi a^3}$ is the dipole coupling coefficient, with $\mu_0$ the permeability of the vacuum, $\mu$ the magnetic moment of each spin. $\vec{S}_j$ is the j-th spin on the lattice, $\vec{r}_{jk}$ is the position of the j-th spin relative to the k-th spin. The sum runs over all pairs of nano-islands (spins). The ground state of the Hamiltonian (\ref{hamiltonian}) is twofold degenerated, one of these configurations is represented on fig. \ref{sjk}(a). Its useful to divide the square spin ice lattice in two sub-lattices, one composed by all the horizontal spins, denoted by $S^x_j$, and the other composed by the vertical ones, $S^y_j$, the sub-lattices spacing denoted by $a$, as depicted on fig. \ref{sjk}(b), each sub-lattice containing $L \times L$ spins. $\vec{S}_j$ denotes spins in both sub-lattices, being $\vec{S}_j = S^x_j \vec{i}$ on the horizontal sub-lattice and $\vec{S}_j = S^y_j \vec{j}$ on the vertical one. $S^x_j$ and $S^y_j$ are thus Ising variables, assuming values equal to $\pm 1$.  The notation we adopt here has the advantage of allowing the reference to each point of confluence of four spins (the vertexes where we can introduce magnetic monopoles) by the index $j$. At the same time, the same index locate the four spins adjacent to this vertex, given by $S^x_j$, $S^y_j$, $S^x_{j-1}$ and $S^y_{j+L}$. 

\begin{figure}[ht]
\begin{center}
\includegraphics[angle=0,width=0.35\columnwidth]{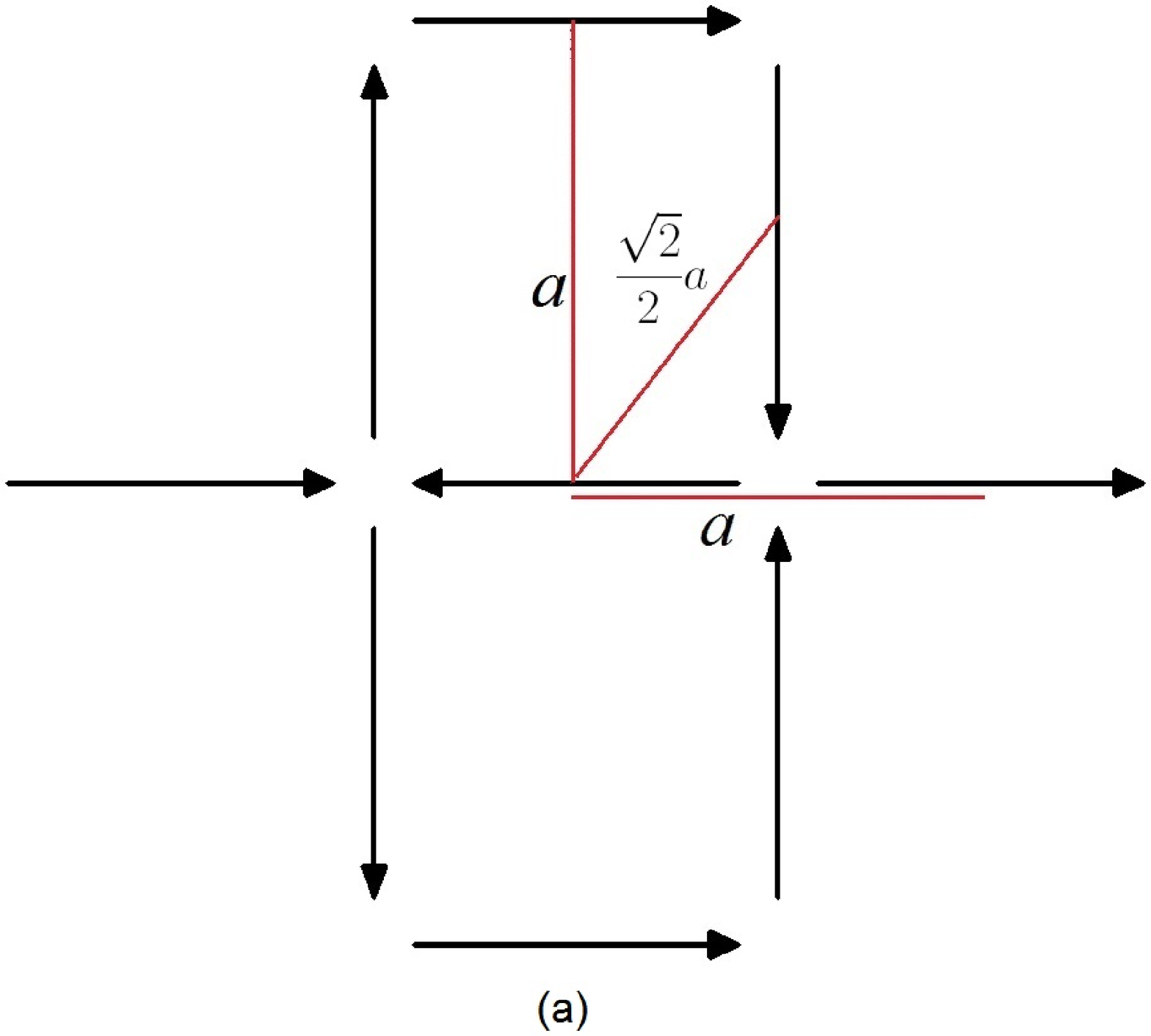} \includegraphics[angle=0,width=0.35\columnwidth]{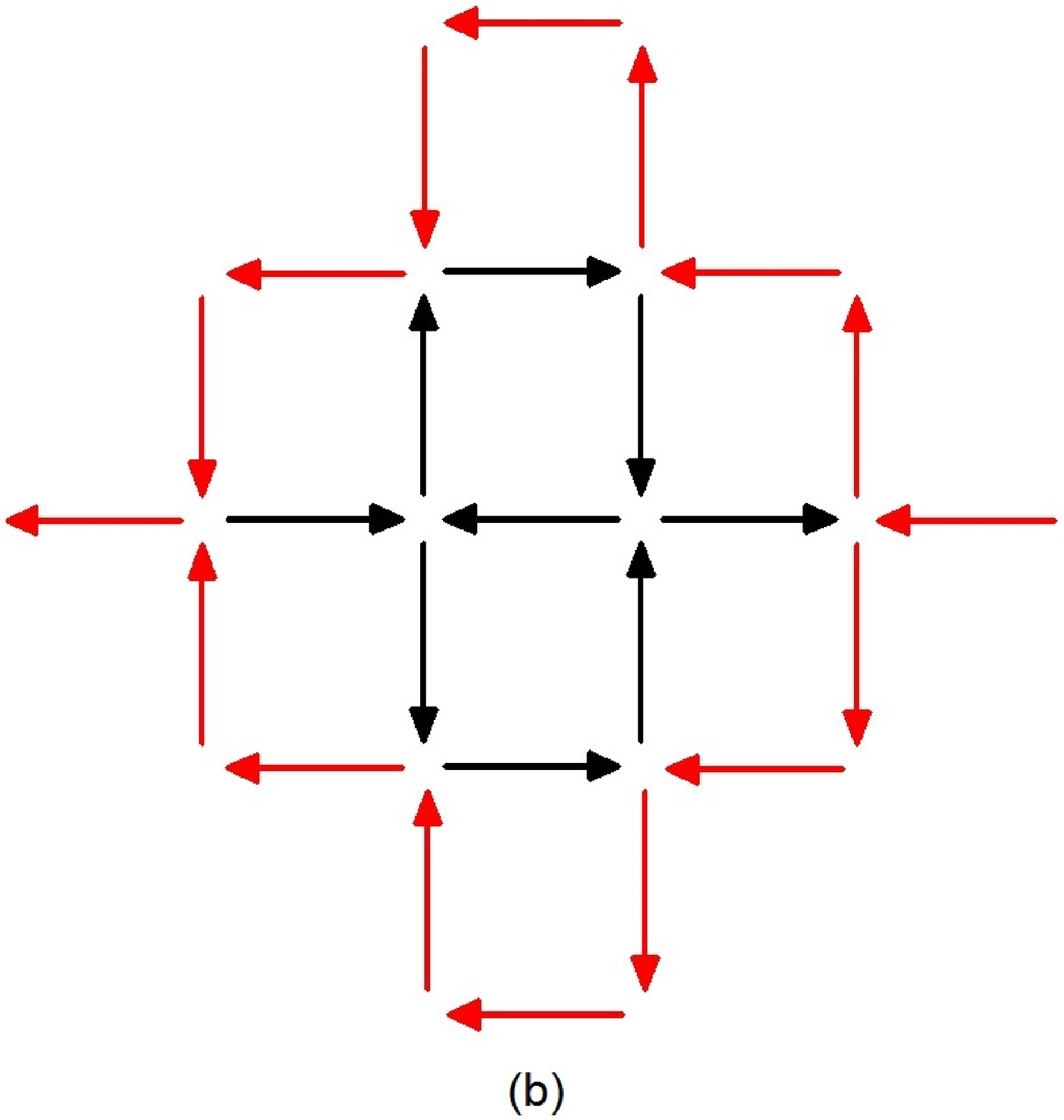}
\end{center}
\caption{(a) The nearest neighbors of a reference spin. (b) The second shell of next-nearest neighbors (in red).}
\label{shells}
\end{figure}

Here we will consider as the nearest neighbors of a given spin $\vec{S}_{j}$ on a sub-lattice its four adjacent neighbors located at both horizontal and vertical directions at a distance $a$ on the same sub-lattice and the other four adjacent neighbors diagonally located at a distance $\frac{\sqrt{2}}{2} a$ on the other sub-lattice, as indicated in figure \ref{shells}(a). This set of eight nearest neighbors surrounds the reference spin as a two dimensional shell, so we will refer to the sets of spins at a given distance of the reference spin as shells. For example, the set of sixteen next-nearest neighbors, indicated in red in figure \ref{shells}(b), will be referred as the second shell of spins and so on. The number of spins on a given shell is equal to $8n$, where $n$ is the index of the shell, and the relative distance $r_{j,k}$ of the spins on the shell to the reference spin grows with $n$, being $r=na$ for the spins located at the edges of the shell.

\begin{figure}[ht]
\begin{center}
\includegraphics[angle=0,width=0.70\columnwidth]{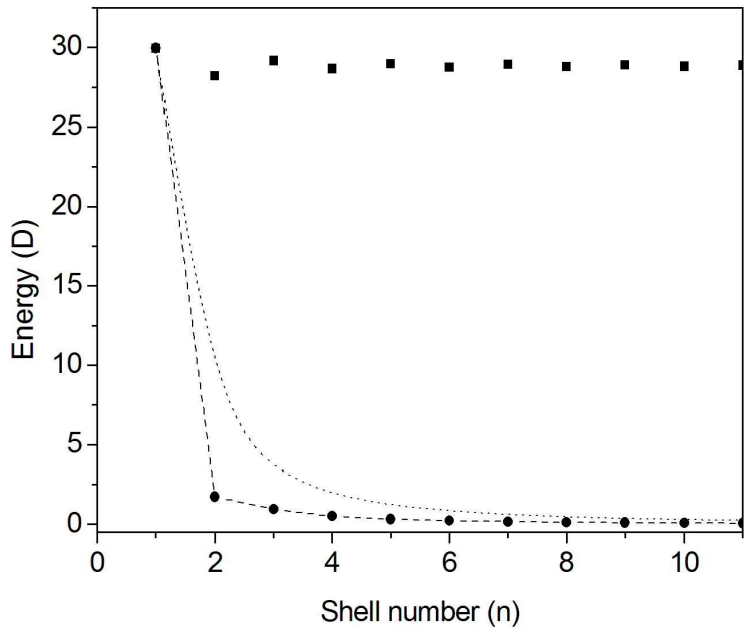}
\end{center}
\caption{The influence of each shell of spins on a reference spin. Full ellipses: the energy cost (absolute values) by flipping the reference spin due to the n-th shell of spins (in units of $D$), dashed line is only a guide for the eyes. Dotted line: the expected $\frac{n}{(na)^3}$ decay if all the contributions of the spins on the n-th shell act on the same direction. The influence of the shells decays faster than this reference decay. Full squares: the cumulative contribution to the energy cost of flipping the reference spin from the first up to the n-th shell.}
\label{figExShell}
\end{figure}

The influence on a specific spin due its eight nearest neighbors relative to the influence of all the other spins on the lattice can be easily computed by means of the following process. Let us compute the energy cost of flipping one single spin, $S^x_{j}$, from the ground state, due to its nearest neighbors. Considering the geometry of the lattice, and computing the contribution to the Hamiltonian (\ref{hamiltonian}) coming from the spin $\vec{S}_{j}$ by means of
\begin{equation}
H_i = D \vec{S}_j \cdot \sum_k \Big( \frac{\vec{S}_k}{r^3_{jk}} - 3 \frac{ \vec{r}_{jk} (\vec{S}_k \cdot \vec{r}_{jk})}{r^5_{jk}}\Big), 
\end{equation}
where the sum in $k$ is taken over the nearest neighbors of $\vec{S}_j$, one can find that this variation of the energy is given by
\begin{equation}
\Delta E_1 = 2D (12 \sqrt{2} - 2), \label{DE1}
\end{equation}
$\Delta E_1 > 0$ since $D > 0$. The same process can be employed to compute now the energy cost relative to the contribution of the next-nearest neighbors, the second shell of spins surrounding $S_{j}$, as shown in figure \ref{shells}(b), resulting in
\begin{equation}
\Delta E_2 = 2D \Big( {\frac{\sqrt{2}}{2}+\frac{1}{4} - \frac{72\sqrt{10}}{125}} \Big), \label{DE2}
\end{equation}
with $\Delta E_2 < 0$. Note the alternating sign of each contribution as we go from one shell to the next one. By proceeding this way for the third shell of spins, as well as for the fourth, fifth and so on, we obtain the result depicted in figure \ref{figExShell}, in which we observe the rapid decay of the energy contribution (in absolute values) of the coupling of each shell with the reference spin to the total energy of the system (closed circles with dashed line). This rapid decay is due to the $1/r^3$ decay law of the dipolar interaction as well as to the alternating sign of the spins on a given shell. If all the contributions of the spins on a given shell acted on the same direction, one should expect that the energy relative to that shell would decrease as $\frac{n}{(na)^3}$, since, as we have seen, the number of spins on the $n$-th shell grows with $n$ and the strength of the interaction decreases with $1/r^3$, with $r \approx na$. The $\frac{n}{(na)^3}$ decay is plotted on fig. \ref{figExShell} as a reference (dotted line), and we can see that the energy cost evaluated here decreases faster than this rate.

Also in fig. \ref{figExShell} we show the cumulative energy cost by flipping the reference spin ($\Delta E$) as we consider the joint influence of a greater number of shells (gray squares). Here, the alternating sign of each shell contribution tends to partially cancel each other as we go further on including new shells, and contributes to the fast convergence of $\Delta E$. We can see that the energy difference by considering only the first shell or up to the 10th shell is very small. In fact, by considering only the first shell, $\Delta E$ is given by Eq. (\ref{DE1}), i.e., $\Delta E \approx 29.9411 D$. By adding the contribution of the second shell, Eq. (\ref{DE2}), the result changes to $\Delta E = \Delta E_1 + \Delta E_2 \approx 28.2124 D$. After including the contributions of the first 20 shells, one obtain $\Delta E \approx 28.8341 D$. As we can see, the contribution coming from the nearest neighbors is very close from the total energy variation computed after including 20 shells. The contribution from the first shell of neighbors (nearest neighbors) is only $3.8\%$ greater than the convergence value, $\Delta E = 28.8435$.

Since the major contribution to the energy change as we flip a given spin comes from its coupling to the nearest neighbors, it seems to be a reasonable approximation to compute only the contributions of these neighbors spins. Of course, better approaches could be constructed by considering the next-nearest neighbors and so on, but with the cost of complexity and increasing computational time. So, the nearest neighbors two dimensional spin ice model (nn2DSI) \cite{Xie2015} can be useful to describe, at least qualitatively, the major features of the two dimensional square spin ice and, as a toy model, the phenomena relative to the pyrochlore three dimensional real spin ice.

Furthermore, we can estimate the errors involved on the nearest neighbors approach in comparison with the long range dipolar interaction model by considering the ratio between the energy costs evaluated above, $\frac{\Delta E_{lr}}{\Delta E_{nn}} = 0.9633$, with $\Delta E_{lr}$ standing for the long range dipolar model and $\Delta E{nn}$ for the nearest neighbors approach. Since $\Delta E$ is expressed in units of the dipolar coupling $D$, the exact energy variation could be recovered from the nearest neighbors result by replacing the dipolar coupling by an effective dipolar coupling $D_{eff} = 0.9633 D$. Thus, we can consider the replacing of $D$ by $D_{eff}$ on the nearest neighbors model results as a first correction to our results.

\section{The model Hamiltonian} \label{nn2DSI}

On the nearest neighbors approach we are dealing with here, each spin at each sub-lattice couples with the four nearest neighbors of the same sub-lattice, with different couplings, however, due to the alignment of the spins and to the structure of the dipolar coupling presented in eq. (\ref{hamiltonian}). So, the coupling between two $S^x$ neighbor spins on the horizontal direction is $-2D$, while the coupling of two $S^x$ spins on the vertical direction is $D$. The contribution to the Hamiltonian of the coupling of a specific spin $j$ on the horizontal sub-lattice with its four nearest neighbors in a $L \times L$ square spin ice lattice is given by
\begin{equation}
H^{xx}_{j} = \frac{D}{2} S^x_{j} (S^x_{j+L} + S^x_{j-L} - 2 S^x_{j+1} - 2 S^x_{j-1}),
\end{equation}
with a equivalent expression for the contribution of an spin on the vertical sub-lattice, $H^{yy}_{j}$. Here, and in what follows, we have already included a factor $\frac{1}{2}$ in order to avoid double counting on the evaluation of $H$. 

Each spin of a given sub-lattice also couples to the four nearest spins of the other sub-lattice, as represented in figure \ref{shells}(a). In our convention we assign the same indexes to a given $S^x$ spin and its nearest $S^y$ neighbor at the lower-left position, as exemplified in figure \ref{sjk}(b). Then, the contribution to the Hamiltonian of the coupling of an specific $j$ spin on the horizontal sub-lattice with its four vertical $S^y$ nearest neighbors is given by
\begin{equation}
H^{xy}_{j} = \frac{3 \sqrt{2} D}{2} S^x_{j} (S^y_{j} - S^y_{j+L} - S^y_{j+1} + S^y_{j+L+1}),
\end{equation}
and the contribution of a spin on the vertical sub-lattice with its four horizontal neighbors is
\begin{equation}
H^{yx}_{j} = \frac{3 \sqrt{2} D}{2} S^y_{j} (S^x_{j-L-1} - S^x_{j-L} - S^x_{j-1} + S^x_{j}).
\end{equation}
Finally, the nearest neighbors Hamiltonian to the artificial two-dimensional square spin ice is given by
\begin{equation}
H=\sum_{j} \Big\{ H^{xx}_{j} + H^{yy}_{j} + H^{xy}_{j} + H^{yx}_{j} \Big\}, \label{Hnn2DSI}
\end{equation}
where the sum runs over  $0 \leq j \leq L^2-1$.

The relative simple structure of the Hamiltonian given by Eq. (\ref{Hnn2DSI}) allow us to use simple and fast algorithms for a Monte Carlo Simulation with periodic boundary conditions, in contrast with Hamiltonian (\ref{hamiltonian}), where the sum runs over all the spin pairs and periodic boundary conditions require the use of the Ewald summation. So, in this approach, we can obtain faster results and/or employ larger lattices, allowing to obtain qualitative results for several observables on the square spin ice as an insightful scenario for studying dynamic and thermodynamic phenomena in both artificial two-dimensional and real three-dimensional spin ice models.

\section{Numerical results} \label{nr}

We simulated the thermodynamics of the Hamiltonian described by Eq. (\ref{Hnn2DSI}) by using Monte Carlo simulation with Metropolis algorithm for different lattices sizes with periodic boundary conditions. As expected, the ground state at zero temperature is twofold degenerated and presents the ice rule structure, as depicted in fig. \ref{sjk}(a). The system exhibits frustration and the mean magnetization is zero in both ordered and disordered phases. So, as the order parameter, we use the pseudo-magnetization, defined as
\begin{equation}
M = \sum_{j,k} (S^x_{j} + S^y_{j}) e^{i\pi f_L(j)}.
\end{equation}
The function $e^{i\pi f_L(j)}$ projects each spin of the lattice on local axes that follow the ice rule structure, $f_L(j)=j$ for an odd $L$, and $f_L = int\Big(\frac{L+1}{L}j\Big)$ if $L$ is even, with $int(x)$ corresponding to the integer part of the argument. The mean pseudo-magnetization per spin is $1$ or $-1$ in the ground states and tends to zero as the system evolves to the disordered phase. 

We use three different lattice sizes, $L=10$, $L=24$ and $L=32$, each one with $2L^2$ spins. As we have already mentioned, due to the nearest neighbors interactions and consequent ease in the implementation of the periodic boundary conditions, the computational procedures are (relatively) very fast. Here, we use typically $10^5$ to $10^6$ Monte Carlo steps, with up to $10$ independent runs, in each simulation. 

\begin{figure}[ht]
\begin{center}
\includegraphics[angle=0,width=0.70\columnwidth]{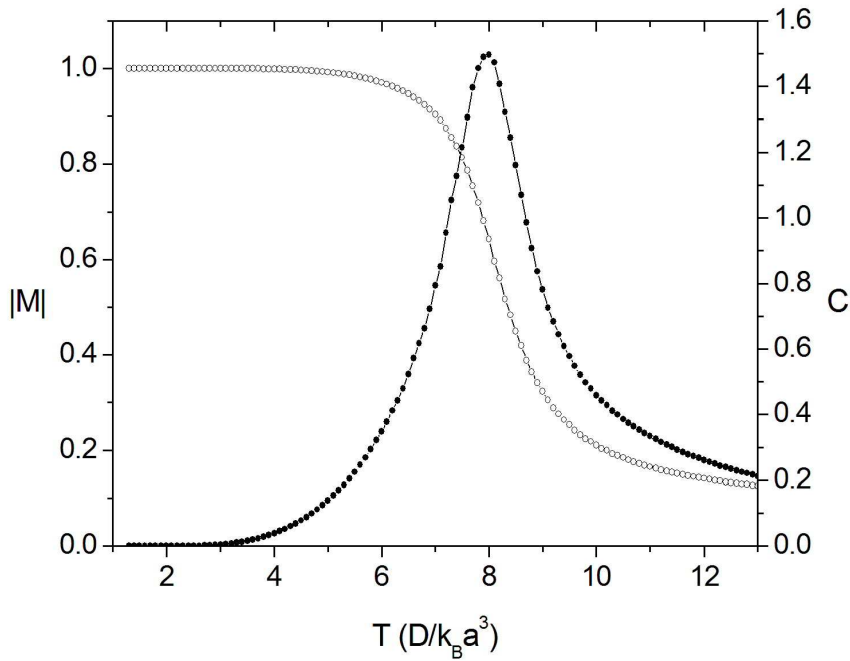}
\end{center}
\caption{Pseudo-magnetization and specific heat for the $L=10$ lattice.}
\label{MCL10}
\end{figure}

\begin{figure}[ht]
\begin{center}
\includegraphics[angle=0,width=0.70\columnwidth]{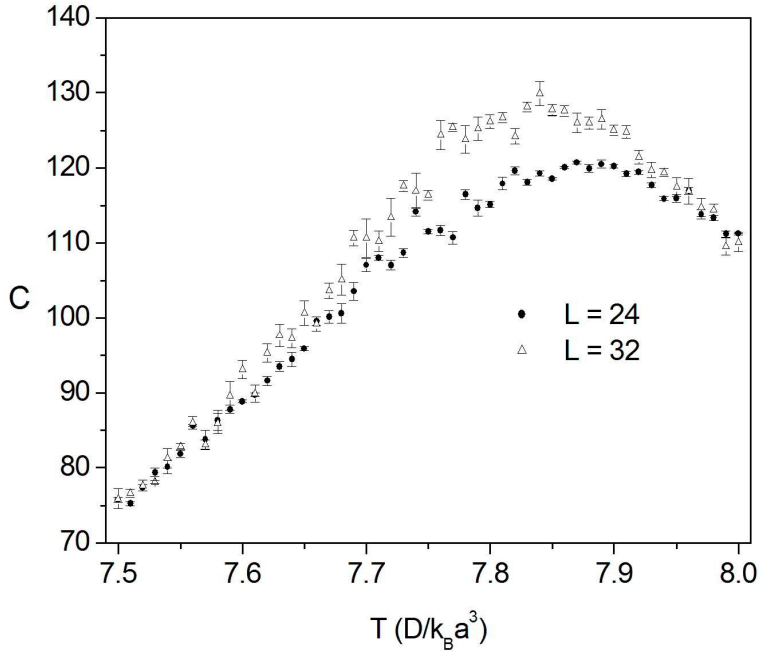}
\end{center}
\caption{Specific heat for the $L=24$ and $L=32$ lattices.}
\label{C24e32}
\end{figure}

In order to determine the critical temperature of the system we observe the absolute value of the pseudo-magnetization as a function of the temperature and also the specific heat of the system. In fig. \ref{MCL10}, we observe the pseudo-magnetization curve for $L=10$, the standard errors bars are smaller than the ellipses height. The specific heat is also presented on fig. \ref{MCL10}, and shows its maximum at $T_C=8.0 D/k_B$. The results for the specific heat with $L=24$ and $L=32$ are shown in figure \ref{C24e32}, where we focused the temperature range near the critical temperatures given by $T= 7.87 D/k_B$ and $T=7.84 D/k_B$ respectively. We observe two main differences relative to the simulations of the long range dipolar interaction system presented in ref. \cite{Silva2012,Li2013} - first, in contrast with the long range results, the critical temperature shows a dependence with the lattice size, as we expect for a nearest neighbor interaction. So, $T_C$ decreases as the size of the lattice grows. Also, the peaks on the specific heat curves are not so sharp as in the long range interaction system. Second, the transition temperature for the nn2DSI system is a bit higher than that for the long range interaction system, as we could expect, since, when we consider other shells besides the nearest neighbors one, the frustrated structure tends to smooth the bond of a specific spin to the rest of the lattice, as we saw on section \ref{sri}. 

\begin{figure}[ht]
\begin{center}
\includegraphics[angle=0,width=0.70\columnwidth]{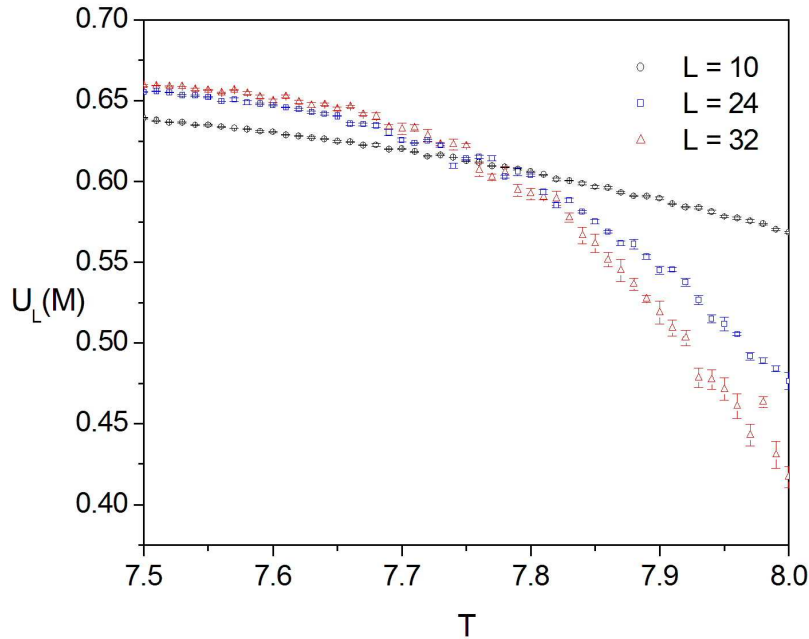}
\end{center}
\caption{The Binder's fourth order cumulant for different lattice sizes. The intersection of the curves points to a critical temperature $T_C = 7.75 D/k_B$.}
\label{figBinder}
\end{figure}

In the context of the finite size scaling theory, by assuming for the nn2DSI model the same universality class of the two dimensional Ising model, as shown in ref. \cite{Xie2015}, we can use the Binder's fourth order cumulant given by
\begin{equation}
U_L(M) = 1 - \frac{<M^4>}{3<M^2>^2}
\end{equation}
to evaluate the $L \rightarrow \infty$ critical temperature from the intersection of the $U_L(M)$ curves for different $L$, as shown on fig. \ref{figBinder}. Within the results we have we can estimate the transition temperature in the range $7.7 \leq T_C k_B/D \leq 7.8$. If we apply the effective dipolar coupling $D_{eff} = 0.9633 D$, as discussed in section \ref{sri}, it points to a transition temperature of $T_C \approx 7.46 D/k_B$, to be compared with the results presented for the long range dipolar model, given by $T_C = 7.2 D/k_B$ \cite{Silva2012} or $T_C = 7.13 D/k_B$ \cite{Li2013}. 

The four spins that are adjacent to a given site $j$ are given by $S^x_j$, $S^y_j$, $S^x_{j-1}$ and $S^y_{j+L}$ (fig \ref{sjk}(b)). Considering the ``{\it flux}'' due to the spins pointing inward (or outward) the vertex $j$ we can define its magnetic monopole charge \cite{Castelnovo2008} as
\begin{equation}
Q_{j} = S^y_{j} + S^x_{j-1} - S^x_{j} - S^y_{j+L}. \label{charge}
\end{equation}
By using eq. (\ref{charge}), we can compute the magnetic charge at each lattice vertex and, consequently, count the number of monopoles with $Q=+2$, $N_2$, or $Q=+4$, $N_4$, at each configuration of the system. We also can distinguish between the string vertexes, those that are part of a string connecting two monopoles (or of a string loop), and the vacuum vertexes. Both present $Q=0$, but in the vacuum the two $S^x$ spins of a given vertex point in opposite directions (the same is valid for the $S^y$ spins). Thus, we can also compute $N_s$, the number of vertexes that have the string structure. We will pay attention here to the single monopoles density, defined by $\rho_2 = \frac{N_2}{N}$, where $N=L^2$ is the total number of vertexes in a $L \times L$ lattice, and the string density, $\rho_s = \frac{N_s}{N}$. The density of monopoles with double charge, $\rho_4 = \frac{N_4}{N}$ showed to be negligible in all range of temperatures we studied here, as already expected, since the energy associated to the double charge monopole vertex is quite higher than the other ones. It is worth to mention that, since $N_2$ counts only the positive monopoles, $\rho_2$ corresponds in fact to the density of pairs of monopoles, occupying each one two lattice vertex, so the maximum $\rho_2$ is $\rho_2 = 0.5$. 

\begin{figure}[ht]
\begin{center}
\includegraphics[angle=0,width=0.70\columnwidth]{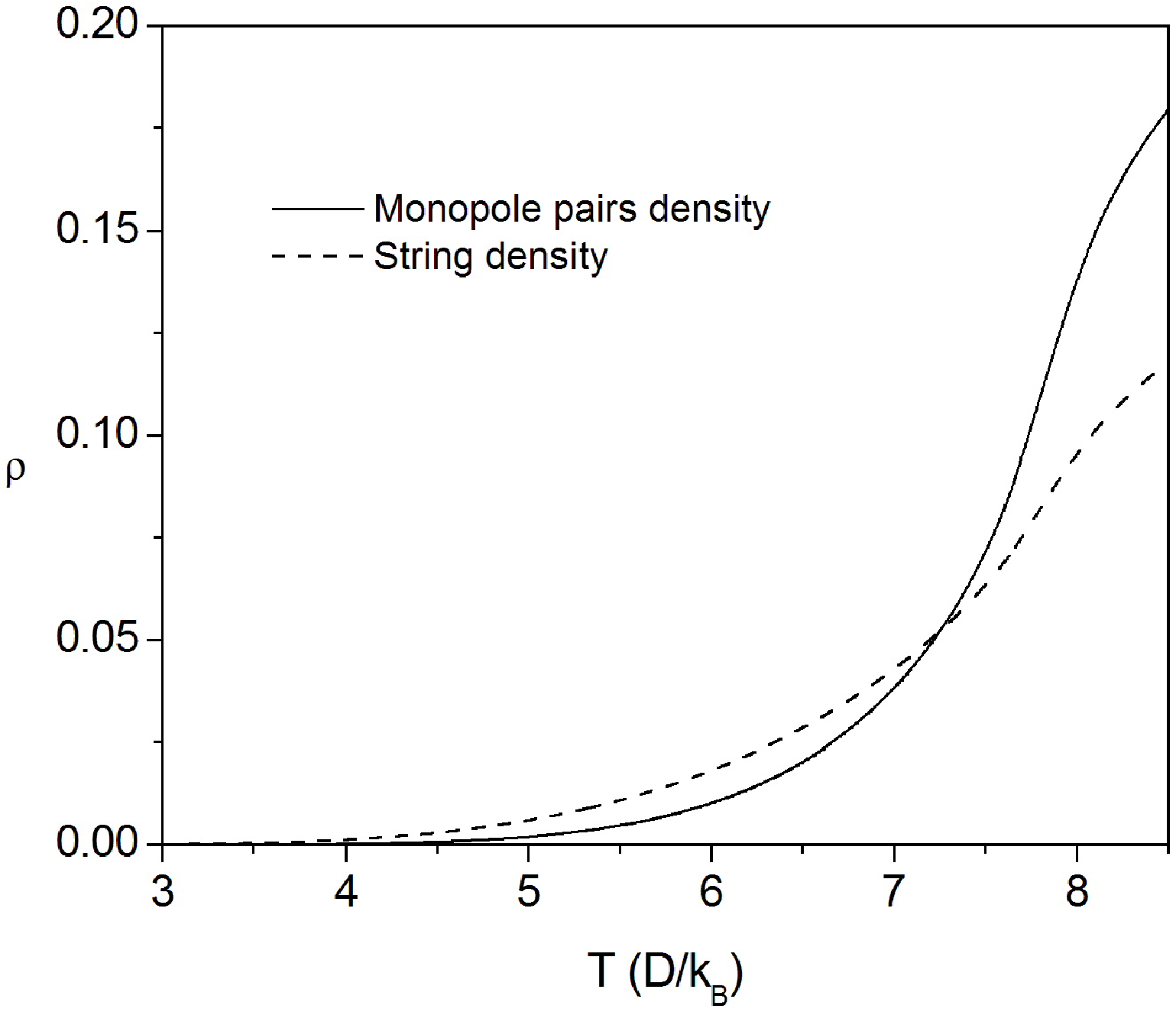}
\end{center}
\caption{The density of monopoles pairs and of string vertexes as a function of the temperature in the nn2DSI model.}
\label{figDensity}
\end{figure}

In fig. \ref{figDensity} we present the behavior of $\rho_2$ and $\rho_s$ as a function of the temperature. As we can see, at low temperatures both monopoles pairs and strings densities are very low, but $\rho_2 \gg \rho_s$, indicating that the states are populated by pairs of adjacent monopoles (dimers). As temperature grows, the density of string vertexes grows, i.e., the states become populated by monopoles separated by strings. At $T_s = 7.3 D/k_B$ we reach the condition $\rho_s = \rho_2$. Above this temperature the states are statistically populated by monopoles separated by strings of length greater than one lattice spacing, at least. Observe that this temperature is lower than the critical one, so there is a range of temperatures where the system is still on the ordered phase but is composed of a gas of separated monopoles. The system smoothly evolves, thus, from a rarefied gas of dimers at low $T$ to a gas of separated monopoles at $T_s$ (although the separation is small) and then to the disordered phase, at $T_c$. At $T_s$ approximately $16\%$ of the lattice vertexes are occupied by monopoles or strings. From this temperature on, both monopoles and strings densities experience a rapid growth, with a occupation of the lattice approximately equal to $25\%$ of the vertexes at $T_c$ and $\rho_s$, approximately twice the monopoles pairs density, $\rho_2$, indicating that the strings lengths are typically two lattice spacing.

\begin{figure}[ht]
\begin{center}
\includegraphics[angle=0,width=0.50\columnwidth]{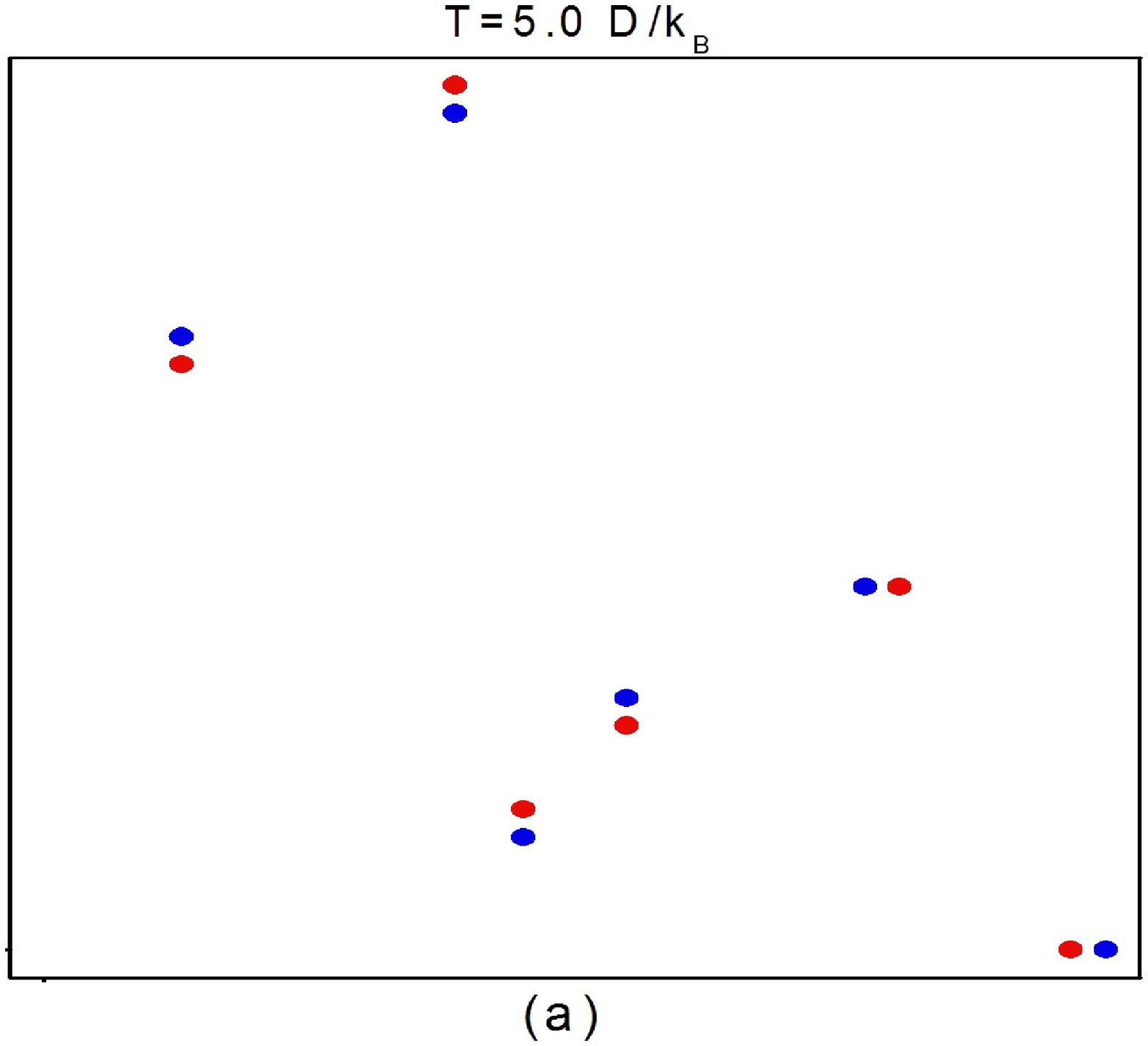}\includegraphics[angle=0,width=0.48\columnwidth]{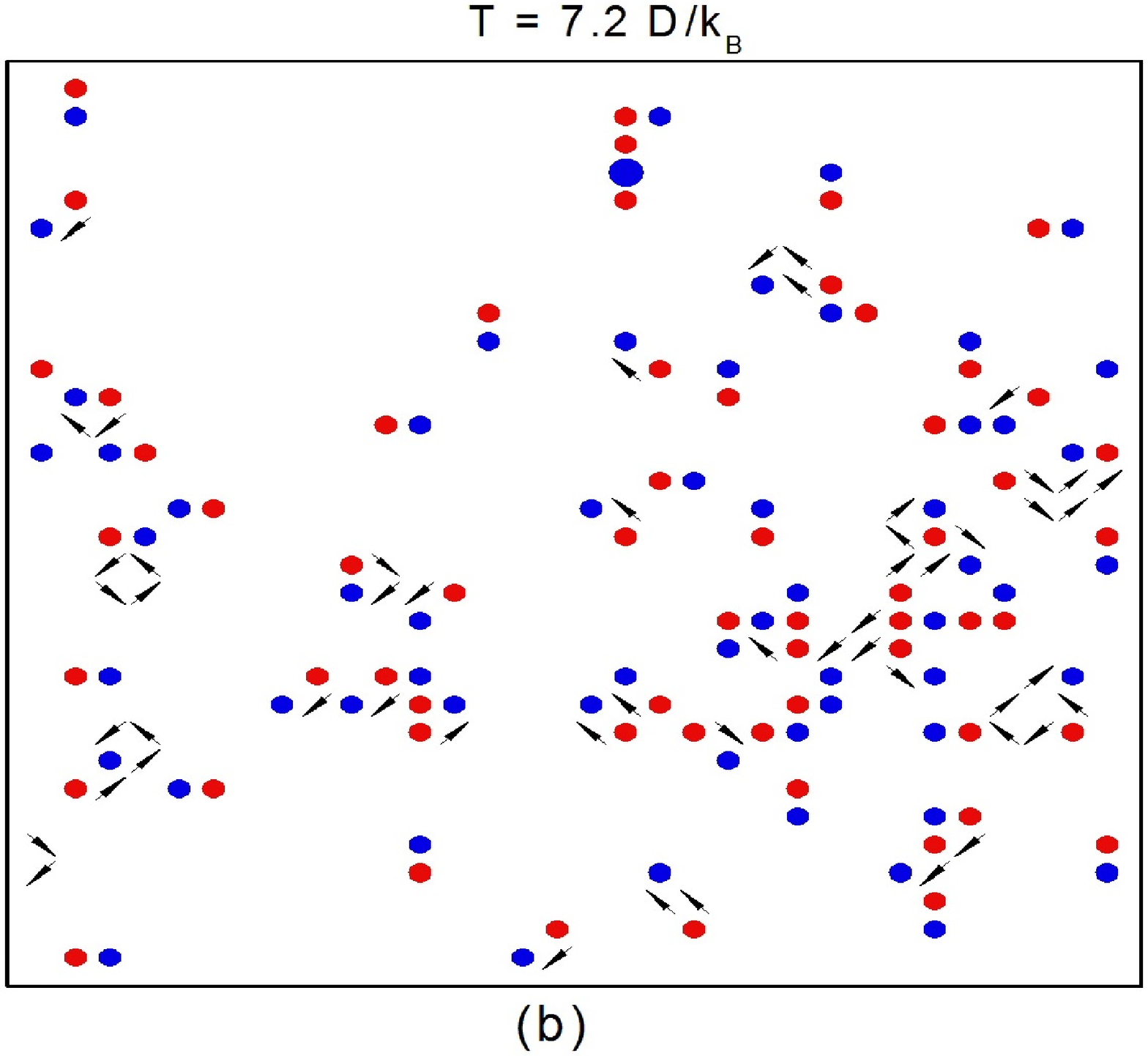}

\includegraphics[angle=0,width=0.50\columnwidth]{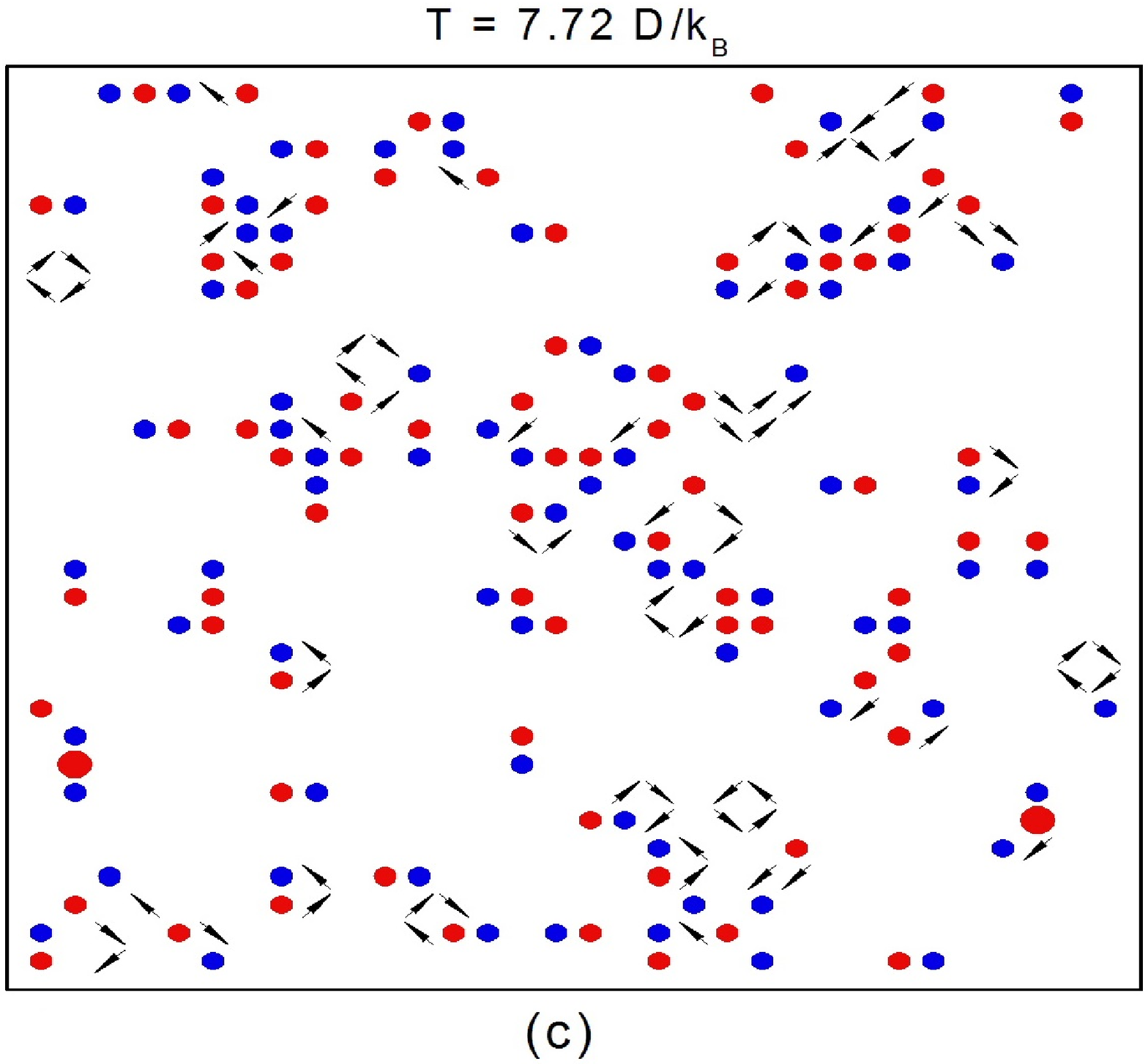}\includegraphics[angle=0,width=0.50\columnwidth]{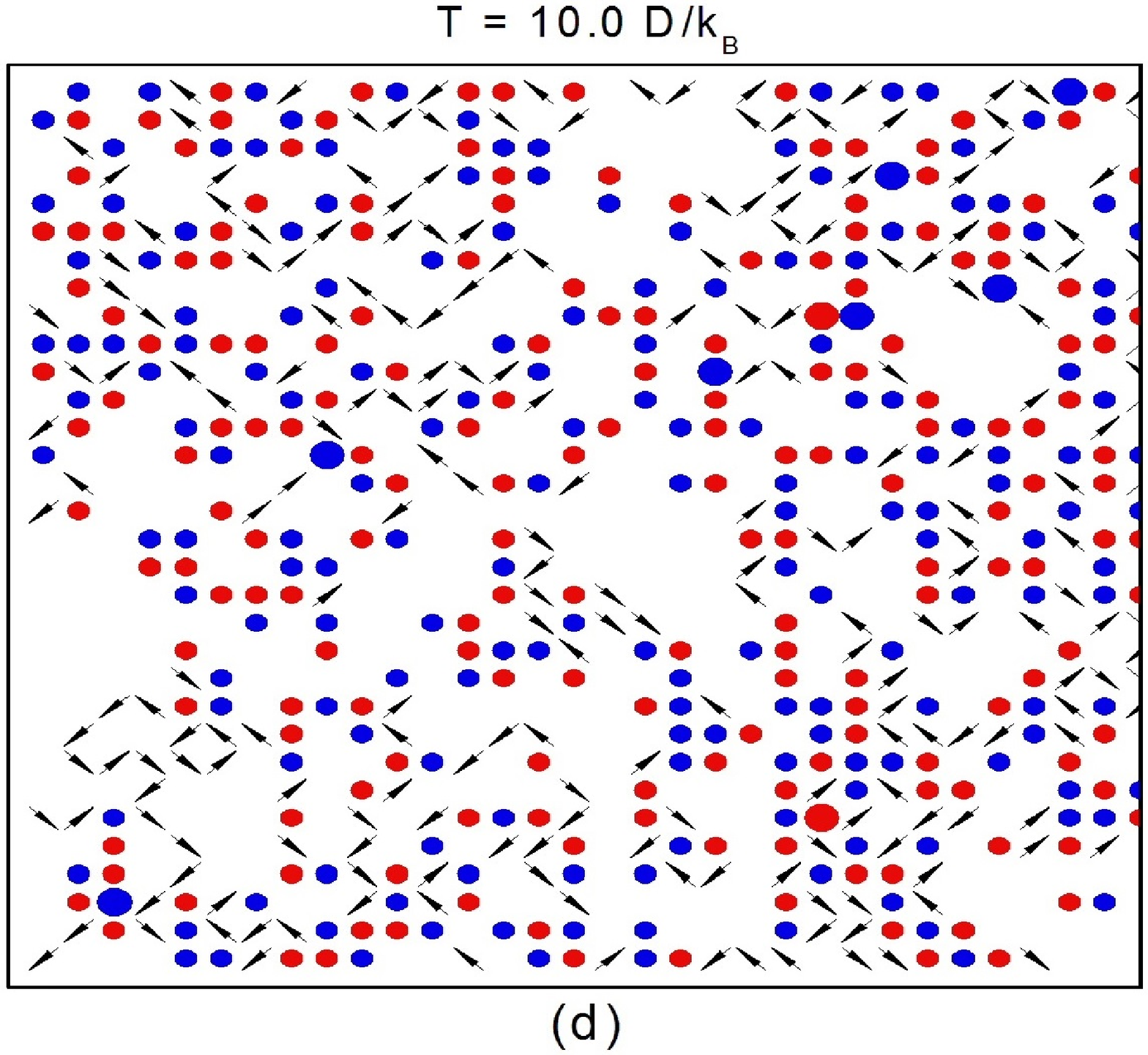}
\end{center}
\caption{Four sampled states at different temperatures on the $L=32$ lattice. (a) At low temperature the states present low densities and are composed by dimers. (b) At the characteristic temperature $T_s$ the density of strings becomes equal to the density of monopoles pairs. (c) Right bellow $T_c$ the strings have mean length approximately equal to two lattice spacings. String loops appear, but their occurrence are still uncommon. (d) In the disordered phase ($T=10 D/k_B$) the states are highly populated by monopoles. Double charged monopoles become more common, the strings are longer and the densities are higher.}
\label{figCharges}
\end{figure}

In fig. \ref{figCharges} we illustrate this evolution by sampling four states from the ensemble at four different temperatures. The red ellipses represent the positive magnetic monopoles (vertexes in which there are three spins pointing inward and one pointing outward the vertex). Blue ellipses represent the negative monopoles. The big red and blue ellipses represent doubly charge monopoles (four spins pointing inward or four spins pointing outward the vertex). The small arrows represent string vertexes, the direction of the arrows is equivalent to the resultant direction for the addition of the four spins adjacent to each vertex. The four snapshots in the figure are sampled configurations,  Fig. \ref{figCharges}(a) shows a state at $T=5.0 D/k_B$, and we observe a low density of dimers. At $T=7.2 D/k_B$ (fig. \ref{figCharges}(b)) the sampled state is already more populated, presenting a reasonably number of monopoles separated by strings, most of them with length of one or two lattice spacing. It is interesting to observe the presence of a string loop (a closed stream of string vertexes, with no magnetic charges), the one positioned at the left/center of the chart. The presence of string loops, however, is not so likely at this temperature. Right bellow the critical temperature, as shown on fig.\ref{figCharges}(c), the number of charges and strings has growth, and a few monopoles with double charge are observed, although its statistical density is still very low. Finally, above $T_C$ the states are more populated, with a greater presence of double charged monopoles and longer strings. 

\section{Conclusion} \label{conclusions}

We have presented here an extension of the nearest neighbor spin ice model for two dimensional square arrays. In comparison with the model proposed in \cite{Xie2015}, this extension includes the coupling with two other spins in such way that the set of nearest neighbors closes a shell around a given reference spin. We computed the influence of the successive shells of spins surrounding the reference spin, and concluded that, at the ground state, the nearest neighbors are responsible for the major influence on that spin, all the rest of an infinite lattice contributing only with $3.8\%$ of the interaction energy. Based on this observation, we define an effective coupling that works as an empirical correction for the $D$ coupling dependent results. We used Monte Carlo simulation to obtain the phase transition of the system and compare the results with the ones obtained by simulation of the long range dipolar interaction, showing a reasonably agreement, justifying the use of the nn2DSI model as a insight model for the study of spin ice phenomena. As an application, we study the evolution of the monopoles and string densities as functions of the temperature, defining the temperature ranges as one can expect the predominance of dimers or separated dipoles.

\section*{Acknowledgments}
The authors would like to thank financial support from CNPq and FAPEMIG (Brazilian funding agencies).

\end{document}